\documentclass[12pt,a4paper]{article}
\usepackage{epsfig}
\usepackage{times}
\usepackage{mathptm}

\makeatletter
\def\fps@figure{tbp}
\def\fps@table{tbp}
\makeatother

\makeatletter
\def\setpicwidth#1#2{\if@twocolumn\setlength{\picwidth}{#2}\else
   \setlength{\picwidth}{#1}\fi}
\makeatother

\def\hc#1{\leavevmode\hbox to \hsize{\hss #1\hss}\leavevmode}
\newdimen\picwidth

\def\epsfigure#1#2{\setpicwidth{#1\textwidth}{\hsize}
  \def\figurefile{#2}
  \hc{\psfig{file=\figurefile,width=\picwidth}}}

\def\super#1{\hbox{$^{\rm #1}$}}

\def\degree{\super{\circ}}

\def\mean#1{\hbox{$\langle #1\rangle$}}
\let\mal\times  

\def\={\penalty10000-\penalty10000\hskip0pt}

\begin{document}

\title{Changes of the cosmic-ray mass composition in the 
   $10^{14}$--$10^{16}$~eV energy range}
   
\author{K. Bernl\"ohr, W. Hofmann, G. Leffers, V. Matheis, \\
        M. Panter, R. Zink \\ 
{\normalsize\em Max-Planck-Institut f\"ur Kernphysik,}\\
{\normalsize\em Postfach 103980, 69029 Heidelberg, Germany}}

\date{December 16, 1997}

\maketitle


\begin{abstract}
{%
Data taken with ten Cosmic Ray Tracking (CRT) detectors and
the HEGRA air-shower array on La Palma, Canary Islands, have been
analysed to investigate changes of the cos\-mic\=ray mass composition
at the `knee' of the cosmic-ray flux spectrum near $10^{15}$~eV
energy. The analysis is based on the angular distributions of
particles in air showers. HEGRA data provided the shower
size, direction, and core position and CRT data the particle track 
information.
It is shown that the angular distribution of muons in air showers
is sensitive to the composition
over a wide range of shower sizes and, thus,
primary cosmic-ray energies with little systematic uncertainties.
Results can be easily expressed in terms of \mean{\ln A} of
primary cosmic rays. In the lower part of the energy range covered,
we have considerable overlap with direct composition
measurements by the JACEE collaboration and find compatible
results in the observed rise of \mean{\ln A}. 
Above about $10^{15}$~eV energy we find no or at most a slow 
further rise of \mean{\ln A}.
Simple cosmic-ray composition models are presented which are
fully consistent with our results as well as the JACEE flux
and composition measurements and the flux measurements of the
Tibet AS$\gamma$ collaboration. Minimal three-parameter composition 
models defined by the same power-law slope of all elements below 
the knee and a common change in slope at a fixed rigidity are 
inconsistent with these data.
}
\end{abstract}


%
%






\section{Introduction}

Cosmic rays from outside the solar system show a smooth flux spectrum
and are almost entirely made up of
protons and fully ionized nuclei. Except for spallation products,
the fluxes of all nuclei follow essentially the same $E^{-2.7}$
power-law in the GeV to TeV energy range, where the primaries
can be identified by direct measurements in space- or balloon-borne
experiments. Only at higher energies the overall cosmic-ray flux
spectrum shows two distinct features near $10^{15}$ and $10^{18.5}$~eV,
as determined by indirect ground-based experiments.
Near $10^{15}$~eV (the {\em knee\/}) the spectrums steepens to about
$E^{-3.0}$ and seems to become flatter again between $10^{18}$ and
$10^{19}$~eV (the {\em ankle\/}).

Despite lack of direct evidence, there is general consensus that the
majority of cosmic rays in the energy range up to the knee 
should be accelerated at the shock waves of
supernova remnants (SNR). For reviews of shock acceleration
see \cite{Blandford-1987} and \cite{Jones-1991}. 
Cosmic rays diffuse out of the Galaxy
on a timescale of $10^7$ years and less.
SNR can easily provide enough power to refurbish the cosmic rays.
Stochastic shock acceleration models and a
transformation of source spectra due to
an energy-dependent diffusion coefficient can explain the
observed power-law spectrum up to $10^{14}$~eV or perhaps even $10^{15}$~eV.
In contrast, there are no generally accepted source models for the
ultra-high energy (UHE) cosmic rays above $10^{15}$~eV.
Unfortunately, the cosmic-ray flux at the knee and higher energies is too
small for direct composition measurements. Therefore, little is
known about the composition in the UHE range. Several results from
ground-based air-shower experiments indicate that the average mass
rises near the knee but also evidence for little change or even
the opposite effect has been presented (for recent reviews
see for example \cite{Khrenov-1993,Wdowczyk-1994,Petrera-1995}).

In recent years, several large air-shower experiments have been
supplemented by additional detector types which enable them
to measure composition\=sensitive quantities on a statistical basis
(not on an event-by-event basis)
and compare these to expected values from shower simulations.
In general the methods are based on different ratios of the
numbers of muons to the numbers of electrons or on the different 
longitudinal development of showers initiated by different primaries.
For such methods both the accuracy of the shower simulations and
of the understanding of detector effects (and how well they can be
included in the simulations) is important. This also applies to
experiments attempting an identification of primaries on an
event-by-event basis.

Results presented in this paper are based on a statistical method
making use of the angular distributions of particles in air showers.
The method has been described in detail in \cite{Bernloehr-1996d}.
Its major advantages are small systematic errors because detector
effects are understood very well and because shower simulations with
different interaction codes give very similar results.
There is also sufficient overlap in energy with direct measurements
to check for any systematic errors.

\begin{figure}
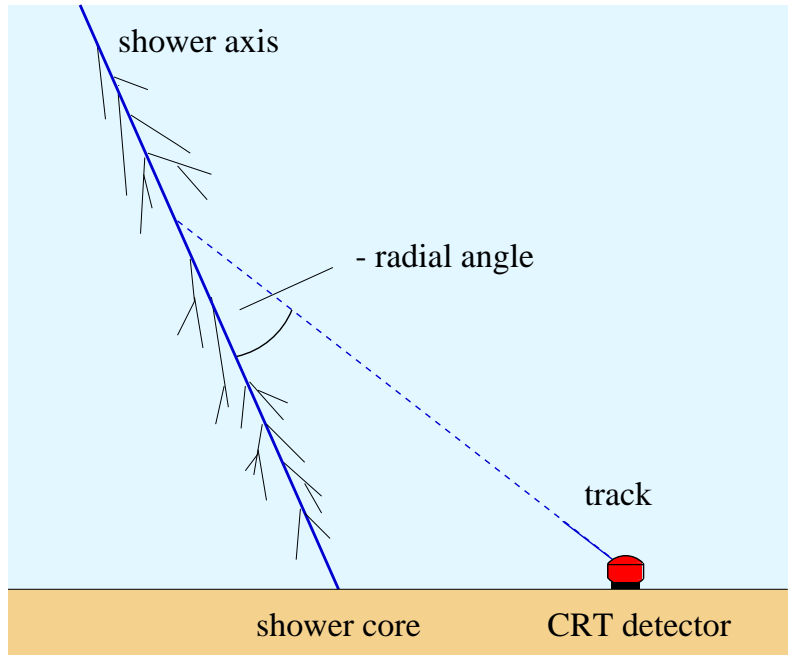

\epsfigure{0.76}{radial-def}
\caption[Definition of radial angles]{Radial angles are defined
as the angles between shower axis and particle track in the
projection into a vertical plane through the core and the
CRT detector location. Note that particle coming from the
shower axis have, in the definition of this work, a negative radial angle.}
\label{fig:radial-def}
\end{figure}

The analysis is mainly based on the angular projections
of shower particles in a reference frame defined by the
positions of the respective tracking detector and the shower
core position. 
The {\em radial plane\/} is defined here as a vertical plane passing
through detector and core positions, and the {\em tangential plane\/}
as a vertical plane perpendicular to the radial plane.
The {\em radial (tangential) angle\/} is the angle between the shower 
and the track projected onto the radial (tangential) plane
(see figure~\ref{fig:radial-def}). Except for vertical showers, the
shower axis is, in general, not exactly in the radial and tangential
planes. Since the analysis was restricted
to showers with zenith angles of less than 30 degrees, this
definition is essentially equivalent to one where radial and tangential
planes are parallel to the shower axis instead of to the vertical
(see \cite{Bernloehr-1996d}).

The measured average radial angles are closely related to the average
longitudinal shower development, with earlier development
(heavier nuclei) corresponding to radial angles closer to zero.
Most muons are produced
near the shower axis and are rarely deflected by more than a few
tenths of a degree due to multiple scattering and the geomagnetic
field. Thus, muon radial angles can, in principle, be transformed into
muon production heights for any given core distance. Since statistical
and systematic errors are much easier to understand directly in terms of
measured angles, no such transformation is used here. Radial angle
distributions are used as a measure of the average longitudinal 
shower development while tangential angles give a measure of our
angular resolution (since the experimental resolution substantially 
exceeds the intrinsic scatter).


\section{Experimental setup and data handling}

The Cosmic Ray Tracking (CRT) project had the initial goal
to build a large air-shower array with tracking detectors
\cite{Heintze-1989b}. Compared to scintillator arrays,
such an array of tracking devices would promise a lower
threshold energy \cite{Bernloehr-1996c} and good angular
resolution even for small shower sizes. To search for gamma-ray
sources, a substantial reduction of the hadronic background
by a very clean muon identification was one of the design goals.

The detectors which emerged out of the CRT project consist of
two 2.5~m\super{2}\ circular drift chambers of the TPC 
(time projection chamber) type on top of each other, 
and a 10~cm thick iron plate as a muon filter
between both chambers \cite{Bernloehr-1996a}. Each chamber has six
readout wires with charge-division and segmented cathode strips
for fully three-dimensional track reconstruction in each chamber.
The whole detector is in a gas-tight container filled with an
argon-methane gas mixture.

Readout is performed through a 40~MHz FADC system into a local
computer system next to each CRT detector. Tasks performed by these
computers include detector control and monitoring but also online
track reconstruction and calibration. Ten prototype
detectors were installed and operated at the site of the
HEGRA air-shower experiment \cite{Fonseca-1995,Aharonian-1996} at the
Roque de los Muchachos Observatory on La Palma, 
Canary Islands for about three years (1993--1996). 
The primary goal of this installation was
to test this new type of cosmic-ray detectors which met all anticipated
design goals. These goals included an angular resolution of 0.4\degree, 
a muon-electron discrimination better than $10^{-3}$,
and robustness at mountain altitude conditions \cite{Bernloehr-1996b}. 
In conjunction with the HEGRA array, the investigation of the 
cosmic-ray mass composition
has been the major scientific application of this installation.

\begin{figure}
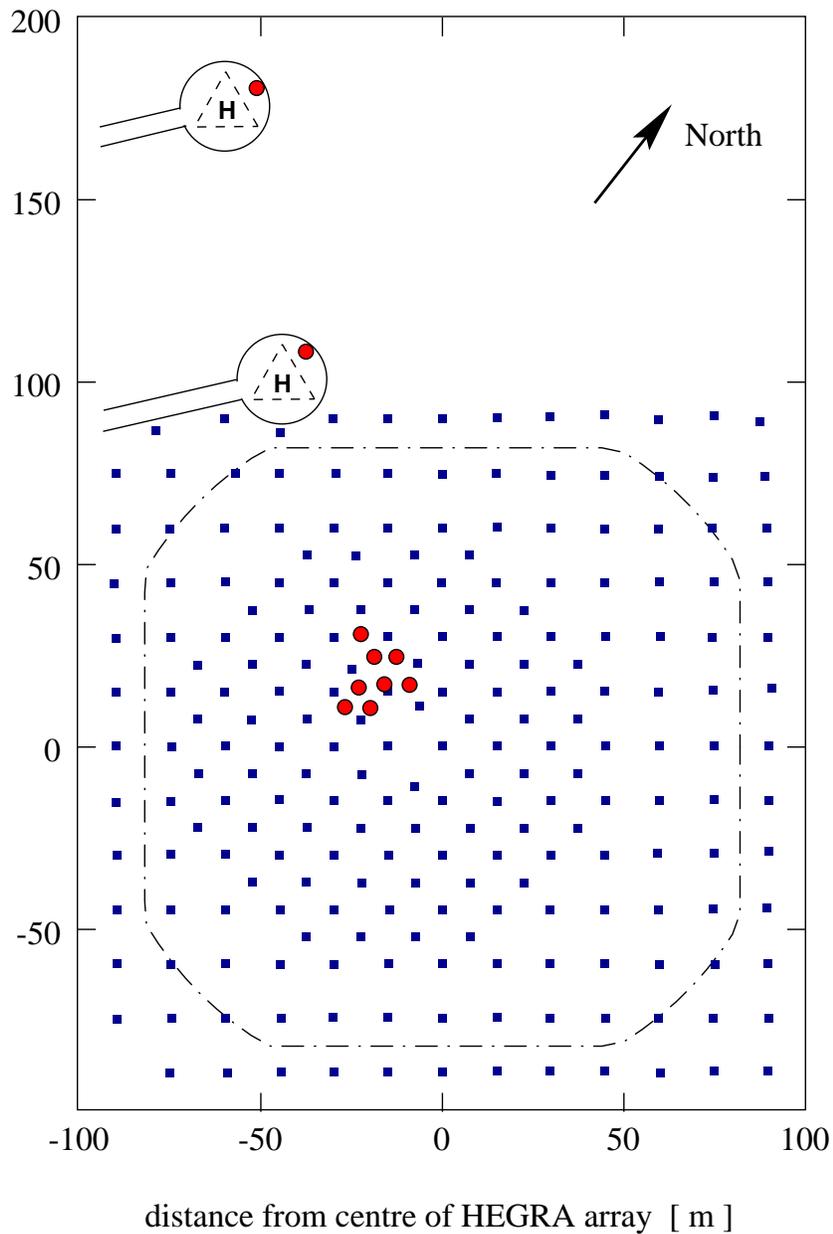

\epsfigure{0.8}{site}
\caption[Map of HEGRA site]{Map of the HEGRA site showing only
the scintillator stations (square dots) and CRTs (circles).
The dash-dotted line marks the area of shower core positions
accepted in the analysis.}
\label{fig:hegra-site}
\end{figure}

The HEGRA experiment \cite{Fonseca-1995,Aharonian-1996} is a 
multi-component air-shower installation with a
detector array and an independent Cherenkov telescope system. 
During the time when the data presented here were taken, the
detector array consisted of 221 scintillator stations of 1~m$^2$
area each (see figure~\ref{fig:hegra-site}), 
49 open photomultipliers with light cones measuring
Cherenkov light (AIROBICC), and 17 detectors of 16~m$^2$ area each with six
layers of Geiger tubes. 
Array data covering part of the time when CRT detectors
were operated with array trigger have been kindly provided by the
HEGRA collaboration. For the independent HEGRA analysis methods and results
see \cite{Aharonian-1996} and references therein.
For the analysis presented here, only
data from the scintillator array have been combined with the
CRT data. Data from the other HEGRA array components have not
been used because these components were in operation only
during a small part of the time when the CRT detectors were operated
with HEGRA triggers and there are not enough combined data
available at energies above about $10^{15}$~eV.

For some of the detector tests and for the composition studies,
the array of CRT detectors was triggered by the HEGRA array
and event numbers were synchronized by the central
CRT electronics system. For the detector tests it was desirable to
have many CRT detectors in a rather small area while for
the composition studies a better configuration 
has detectors distributed over a larger area.
As a compromise, two CRT detectors were installed outside of
the HEGRA array on two existing helicopter ports
(see figure~\ref{fig:hegra-site}). With this configuration
it was possible to measure shower particles at core distances
up to 300~m. For about half of the time covered in the
analysis all ten CRT detectors were triggered by HEGRA.
During the rest of the time the eight detectors inside the
HEGRA array were operated independently of HEGRA for detector tests
while the two on the helicopter ports were taking data with
HEGRA triggers.

The HEGRA and CRT arrays were only loosely coupled, i.e.\ 
triggers, run and event numbers, and Rubidium clock information 
was exchanged but data were re\-corded separately.
Initially, the CRT detectors recorded partially processed hit data 
(pulse areas and drift times). 
The track reconstruction and calibration for these
data were done offline. After the relevant algorithms had been
checked \cite{Bernloehr-1996b} the complete reconstruction and
calibration was done online by the computers at the CRT detectors.
For the HEGRA scintillator array, the ADC and TDC data for all
triggered stations were recorded. Shower reconstruction,
i.e.\ reconstruction of the arrival direction, the position of
the shower core, and the shower size, was done offline.
HEGRA data were made available after this data reduction and
merged with the corresponding CRT data. The available amount of data
for this analysis corresponds to a total life time of 42 days, 
scattered over a period of about one and a half years. 


\section{Simulations}

In extensive air showers where particles are observed after several
interaction lengths of the particle cascade, any quantitative interpretation
of data requires comparison with simulations of physical processes in
the showers as well as in the detectors.
Shower simulations are based on the CORSIKA \cite{CORSIKA-1992} 
program in versions 4.068 and 4.50\@. 
Simulations with different interaction models
were used to obtain an estimate of systematic errors due
to differences in interaction models. 
As the more sophisticated and, therefore,
presumably more accurate model we used the CORSIKA option with
VENUS \cite{Werner-1993} above 80~GeV energy and GHEISHA \cite{Fesefeldt-1985} 
at lower energies.
Completely different interaction models are provided by the
DPM (dual parton model inspired) and the Isobar interaction models
above and below 80~GeV, respectively. For some of the simulations,
the showers were fully simulated, including the electromagnetic
subshowers. All simulations were done for the HEGRA site on La Palma.

As far as the comparison with muon angular distributions
are concerned, it turned out that for showers well above 
the HEGRA trigger threshold it is sufficient to take the shower sizes
from the summed analytical estimates of the electromagnetic subshowers.
Without full simulation of the electromagnetic part a much larger
number of showers could be simulated without exceeding manageable
amounts of disk and tape storage. In total about 30000 proton
and about 15000 iron showers have been simulated, most of them
in the energy range from 20~TeV to 10~PeV for protons and from
40~TeV to 20~PeV for iron. Some simulations extend down to lower
energies. The zenith angles of simulated showers extend up to 32\degree.

Simulations have also been done for showers initiated by
helium, nitrogen (representing the CNO group), and 
magnesium nuclei (representing Ne to S) in order to evaluate
realistic mixed compositions like those measured by JACEE up to
a few hundred TeV \cite{Asakimori-1993ab,Tominaga-1995}. 
All simulations were done with an $E^{-1.7}$ differential
flux spectrum and appropriate event weights (e.g. $\propto${}$E^{-1}$)
were applied to match a desired flux spectrum. 

Shower simulations are followed by simulations of tracking
detectors and the array. The simulation of the tracking detectors 
is quite detailed in the case of muons where multiple scattering 
and energy loss in the iron plate and their consequences on the 
detection efficiency and the reconstructed track angles are fully 
taken into account. Other efficiencies, like those of the identification
of electrons, gammas, protons, or pions as either electron tracks
(in the upper drift chamber only) or muons (pairs of
isolated tracks in both chambers matching in angle to better than
2.5\degree\ in both projections) and the relevant angular
resolutions were parametrized on the basis of detailed detector simulations
with the GEANT \cite{GEANT-manual} package. 

The array simulation used is relatively simple and, thus,
applied only at shower sizes where HEGRA is fully efficient for
showers with cores inside the array. Detailed simulations show that this is
the case for shower sizes $N_{\rm e}$ above 15000. 
For sizes between 10000 and 15000 HEGRA should be already 95\% efficient.

Because the HEGRA scintillation counters are covered by 5~mm of lead,
the measured shower sizes $N_{\rm h}$ do not represent the numbers
of charged particles above the lead but below. A fit to Monte Carlo
simulations \cite{Krawczynski-1995} shows that, on average, a relation
\begin{equation}
N_{\rm h}/N_{\rm e} = (1.7\pm0.2)\,(N_{\rm e}/10000)^{-0.14\pm0.04}
\label{eq:Nh/Ne}
\end{equation}
holds at least in the range $5\mal10^3<N_h<10^6$.
This is independently confirmed by comparing the
experimental relation between the number of fired scintillation counters
and the reconstructed $N_h$ with the simulated relation between
the number of scintillation counters and the $N_{\rm e}$ number
given by CORSIKA. 
The statistical accuracy of $N_{\rm h}$ reconstruction 
is also obtained from simulations \cite{Martinez-1995} and is included
as an additional source of shower size fluctuations.

The angular resolution and the accuracy of the core
location are parametrized as functions of HEGRA shower size and 
zenith angle. 
The resolutions of shower size and core location are taken from detailed
array simulations \cite{Krawczynski-1995,Martinez-1995}.
Their roles are relatively uncritical for the analysis presented in the
following section. For the array angular resolution see also the next section.

After folding in the HEGRA angular, core, and size resolutions, the
same cuts as to measured data were applied and all tracks were
filled in histograms of identical bin sizes as with the measured data.
Each track was given a weight corresponding to the probability that
a particle (muon, electron, proton, pion, etc.)\ would be identified
as a muon times the event weight which corrects from the
$E^{-1.7}$ differential spectrum of simulated showers to the
assumed actual spectrum with or without a knee.


\section{Analysis}

The tangential angles of muons have a very narrow intrinsic distribution
(see \cite{Bernloehr-1996d}) and are used to derive the combined angular
resolution of CRT tracks and HEGRA showers as functions of
shower size, zenith angle and core distance. The HEGRA angular
resolution assumed in the simulations was chosen such that the measured
width of the tangential angle distribution was reproduced in
all cases.

The radial angles of muons but also of electrons
are related to the longitudinal shower development and
are sensitive to the composition.
With the 25~m$^2$ total area of the ten CRT detectors 
there is rarely more than one muon and only a few
electrons recorded in a typical event, and statements on an
event-by-event basis are not possible. Therefore,
this analysis is based on the {\em inclusive} radial angle
distributions and their comparison with simulations.
In particular, we use the median radial angle as a function
of core distance in different intervals of the shower size.

The shower parameters, i.e.\ core position, reconstructed
size $N_h$, and direction are obtained from the HEGRA shower
reconstruction \cite{Krawczynski-1995,Krawczynski-1996}. 
For about 10\% of the
data both the scintillator array and the AIROBICC array of open
Cherenkov counters was operational. For this part of the data,
the shower direction is available separately from both arrays.
It turned out that median radial angles of muon tracks with respect
the AIROBICC shower direction follow the same curve (as a function
of core distance) for all CRT detectors. With respect to the
scintillator array, shower direction deviations of up to a few tenths of
a degree became apparent, in particular for core positions near the
edge of the array. This seems to be due to the fact that the particles 
in the shower are much more concentrated near the shower axis than the
Cherenkov light and the scintillator array is thus more sensitive to
errors in cable delay calibrations. To correct for that effect
after the shower reconstruction, a correction depending on the
core position was applied to scintillator array shower directions.
The amount of this correction is just the average difference between
shower directions from scintillator array and AIROBICC and is slightly
different for data from different periods.
By applying the correction we could take advantage of the smaller systematic
errors of the AIROBICC shower direction but preserve the ten-fold larger
statistics of the scintillator data.

Another calibration aspect which has been very carefully checked
is that of the alignment of the CRT detectors. Their alignment with
respect to the vertical direction was once measured with a large level
mounted on top of the detector containers. Tolerances of the drift
chambers inside the containers are of the order of 1~mm on a 2~m radius.
All containers were measured before assembly. 
Changes of the alignment were monitored
with built-in clinometers with 0.01\degree\ resolution and no changes
were found over a two-year period. For the detectors inside the
HEGRA array independent measurements of the alignment were obtained
from the average angles between tracks and showers for shower cores
near (e.g.\ less than 30~m from) each CRT detector. Both calibrations
agree with a r.m.s.\ scatter of 0.12\degree\ and mean differences
consistent with 0\degree$\pm$0.05\degree. 

For the two detectors
outside of the array the alignment was checked by triggering them
with vertical muons. A purpose-built device with 
two 5~cm diameter scintillators and photomultipliers in a one meter
long tube was mounted
and vertically aligned on top of the detectors at the experimental site.
Good agreement was also found in that case. Errors in the alignment
with respect to the HEGRA zenith direction
are estimated to be 0.05\degree\ for the detectors inside the HEGRA
array and 0.10\degree\ for the two detectors outside.
Calibration of the azimuthal alignment of CRT detectors with respect
to HEGRA is based on the fact that the measured distribution
of tangential angles has to be
symmetric and centered at zero. 
This calibration is accurate to 0.02\degree.


\section{Results}

\begin{figure}
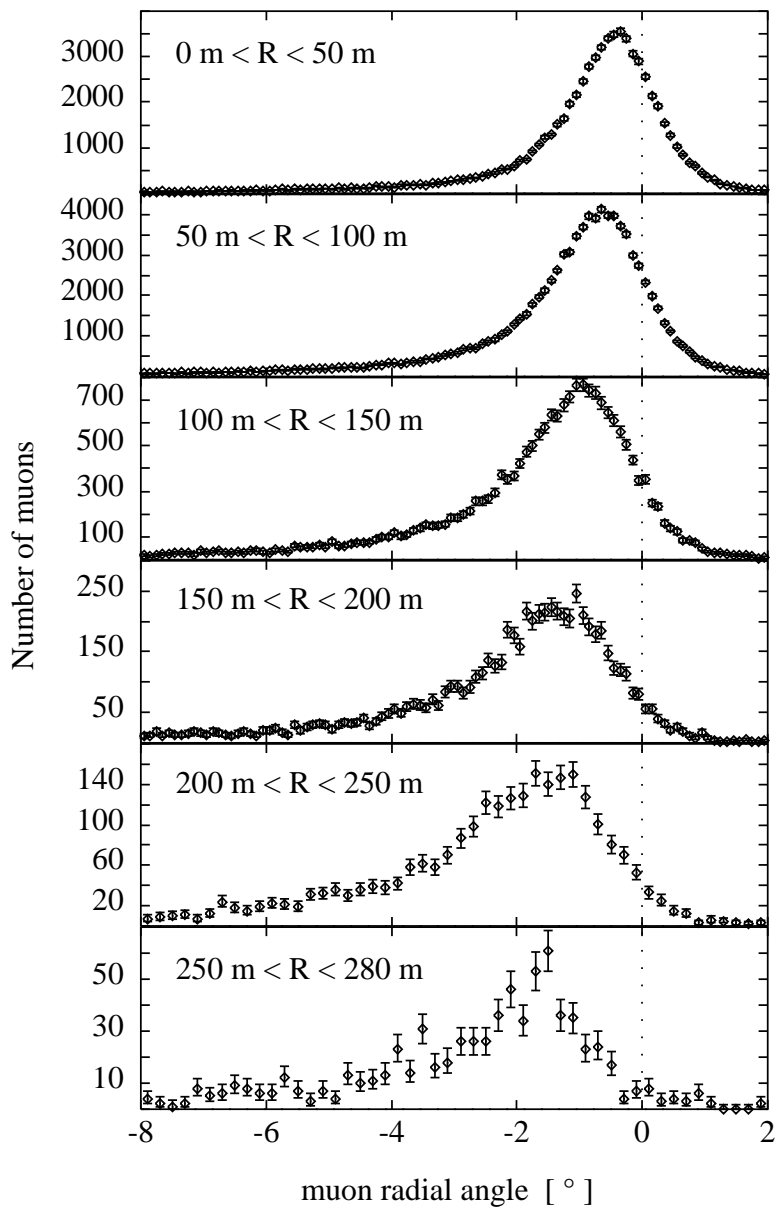

\epsfigure{0.75}{radial-distrib}
\caption[Muon radial angle histograms]{Histograms of muon radial
angles in different intervals of core distances for
$15000<N_{\rm e}<50000$ ($24000<N_{\rm h}<67000$). Note that
particles coming from the shower axis have negative radial angles.}
\label{fig:mu_radial_bins}
\end{figure}

Figure \ref{fig:mu_radial_bins} shows histograms of muon radial angles for
different core distance intervals. Because the distributions have
long tails towards large negative radial angles 
(in the definition used throughout this work, see 
figure~\ref{fig:radial-def}), their median values
have smaller relative statistical errors than their mean values --
opposite to a Gaussian where the median is $\sqrt{\pi/2}$ times worse
than the mean. For the comparison with simulations, the median
values are also more robust against non-Gaussian tails of the
assumed HEGRA angular resolution function and the small fraction
of non-shower muons. The statistical errors of the measured
median values are derived as a given fraction of the
r.m.s.\ deviation of radial angles. The appropriate fraction is
taken from simulations given the expected
shapes of the distributions. Although the central limit theorem
applies to these medians, it should be kept in mind that
probability distributions of the medians are not Gaussian for
very small numbers of muons (but still better than distributions
of the mean). However, this plays only a role in the largest shower size
intervals at large core distances.

\begin{figure}
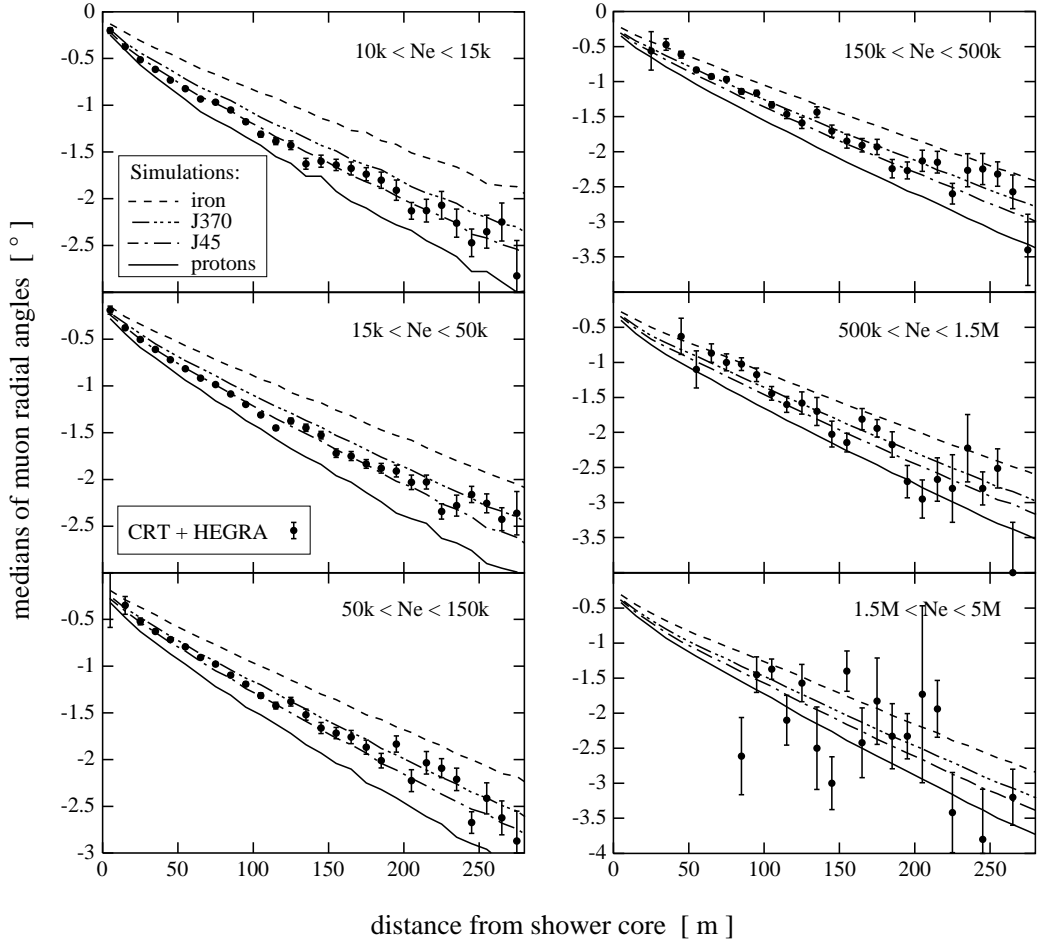

\epsfigure{1.0}{781a-786b-bw}
\caption[Median muon radial angles]{Median radial angles of muons
measured with CRT detectors versus the HEGRA shower axis as functions
of horizontal distance of the CRT detectors from the shower cores.
Simulations are superimposed for pure protons and iron as well as
for mixed compositions (assumed energy-independent with an
$E^{-2.7}$ spectrum)
corresponding to direct measurements of
the JACEE collaboration \cite{Asakimori-1993ab} above 45 TeV
(J45) and above 370 TeV (J370).
Simulations shown are based on CORSIKA interaction options VENUS and
GHEISHA.}
\label{fig:781-786}
\end{figure}

Resulting measurements of median muon radial angles in
10~m core distance bins and for six different intervals of
shower sizes are shown in figure~\ref{fig:781-786}.
Energies corresponding to these ranges are summarized in
table~\ref{tab:E-in-ranges}. At shower sizes below about $10^5$ 
which overlap with direct measurements we find very good agreement 
with the corresponding simulations. Note that for large shower sizes
the CRT detectors cannot reconstruct muons near the shower core,
mainly because the readout electronics is saturated by the signals
of many particles \cite{Bernloehr-1996b}.

\begin{table}
\caption{Average energies (PeV) of simulated showers in six different shower
size intervals, assuming an $E^{-2.7}$ spectrum.}
\label{tab:E-in-ranges}
\begin{center}
\begin{tabular}{crrrrr}
\noalign{\vspace{4pt}}
$N_e$ & \multicolumn{1}{c}{protons} & 
 \multicolumn{1}{c}{He} & \multicolumn{1}{c}{N} & 
\multicolumn{1}{c}{Mg} & \multicolumn{1}{c}{Fe} \\
\noalign{\vspace{4pt}}
\hline
\noalign{\vspace{4pt}}
(1.0--1.5)\,$\mal10^4$ &  0.057 & 0.086 & 0.12 & 0.15 &  0.19 \\
(1.5--5.0)\,$\mal10^4$ &  0.105 & 0.15 & 0.22 & 0.25 &  0.32 \\
(0.5--1.5)\,$\mal10^5$ &  0.29 & 0.41 & 0.54 & 0.60 &  0.78 \\
(1.5--5.0)\,$\mal10^5$ &  0.77 & 1.0 & 1.3 & 1.5 &  1.8 \\
(0.5--1.5)\,$\mal10^6$ &  2.3 & 2.8 & 3.7 & 3.9 &  4.7 \\
(1.5--5.0)\,$\mal10^6$ &  5.7 & 7.2 & 8.8 & 9.2 & 10.5 \\
\noalign{\vspace{4pt}}
\hline
\end{tabular}
\end{center}
\end{table}

Neglecting non-Gaussian tails of the median angles, 
a measure of the cosmic-ray mass composition can be obtained from
the average position of measured data between proton and iron
simulations:
\begin{equation}
\Lambda = \frac{1}{\sum w_i}\sum_i \,w_i\,\frac{\langle\alpha_i\rangle -
  \langle\alpha_{i,{\rm p}}\rangle}{\langle\alpha_{i,{\rm Fe}}\rangle -
  \langle\alpha_{i,{\rm p}}\rangle},
\label{eq:Lambda}
\end{equation}
where $\langle\alpha_i\rangle$ is the measured median muon radial angle
in core distance interval $i$ and $\langle\alpha_{i,{\rm p}}\rangle$
and $\langle\alpha_{i,{\rm Fe}}\rangle$ are from simulation
for pure protons and pure iron nuclei, respectively. 
The weights $w_i$ take the statistical accuracy of measured data
into account. 
By definition, $\Lambda=0$ if the data match
simulated protons and $\Lambda=1$ if they match simulated iron.

\begin{figure}
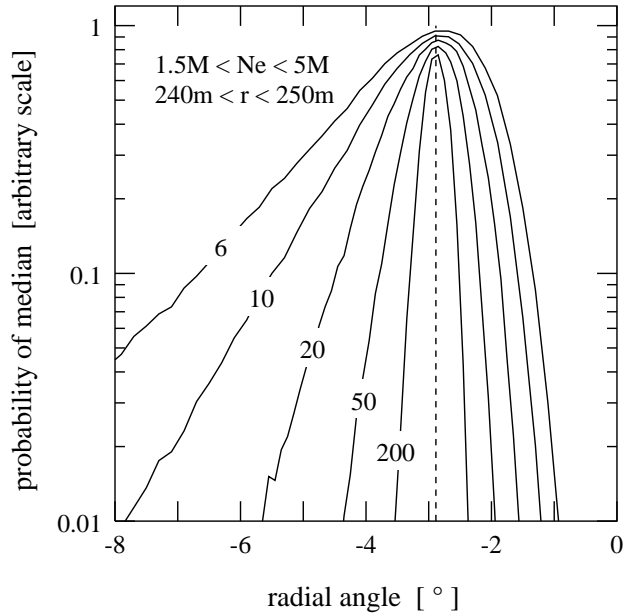

\epsfigure{0.6}{median-distrib}
\caption[Likelihood distributions of medians]{Likelihood distributions
(arbitrarily scaled) for median radial angles with 
6, 10, 20, 50, and 200 entries in
the histogram, based on simulations for one core distance interval 
(240--250~m) in the largest shower size interval with a J370 composition.
Note that the distributions are asymmetric for small numbers of
muons but approach a Gaussian with increasing numbers.}
\label{fig:median-distrib}
\end{figure}

Least squares fits of $\Lambda$ times the iron curve plus
$(1-\Lambda)$ times the proton curve to the data are equivalent to
equation~\ref{eq:Lambda}. Due to the
non-Gaussian tails in case of poor statistics, 
more appropriate maximum-likelihood (ML) fits with $\Lambda$
as the only free parameter were actually used. 
For the ML fits a set of parametrized likelihood
functions was obtained from a range of simulated angular distributions
like those shown in figure~\ref{fig:median-distrib}. For that
purpose, $n$ radial angles ($5<n<1000$) were randomly selected many times
from the simulations. Histograms of the corresponding
median radial angles were stored (one histogram for each core distance
interval, each $n$, and
each element or element mixture). In a second step a suitable parametrisation
was fitted to these histograms and used for the actual ML fits as a third step.
The parametrisation takes into account that, for small $n$, the
distributions of medians are, for technical reasons, slightly different
for odd and even $n$.
As the statistics improves ($n$ grows), error distributions of the median
radial angle values indeed approach a Gaussian and therefore least squares and
ML fits become equivalent. Except for the two largest shower size
intervals, results from both types of fits in fact differ by less than
0.005 in $\Lambda$ and its error. 

\begin{figure}
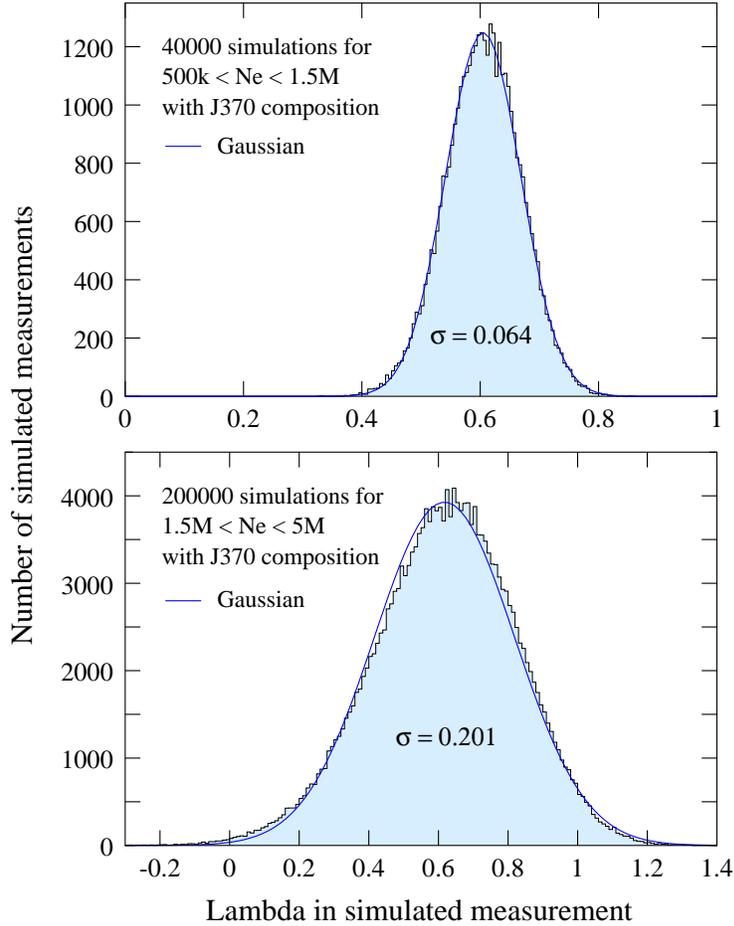

\epsfigure{0.7}{lambda_prob}
\caption[Likelihood distributions of $\Lambda$]{Histograms of resulting
$\Lambda$ values in large numbers of simulated measurements for
shower size intervals $5\mal10^5<N_e<1.5\mal10^6$ (top) and
$1.5\mal10^6<N_e<5\mal10^6$ (bottom). In each core distance interval
the same number of muons as in the experimental data was randomly selected
from simulations for a J370 composition and the same procedures for 
calculating medians and the same maximum-likelihood fits were applied
as to experimental data. The $\sigma$ values of the fitted
Gaussians correspond to the statistical errors of the measurements.}
\label{fig:lambda-prob}
\end{figure}

By simulations, the ML fits were checked to have
no significant bias for our data even if including core distance bins with
only six muons. Such simulations -- using one set of simulated angular
distributions and the experimentally measured numbers of muons
in each distance bin -- were also used to verify the experimental
statistical errors of our $\Lambda$ values 
(see figure~\ref{fig:lambda-prob}). That procedure also showed that
in the largest shower size interval a least squares fit is less accurate
than the applied maximum likelihood fit.
Statistical errors of the air-shower simulations were estimated by
comparing results obtained with separate subsets of the
simulated showers.

\begin{figure}
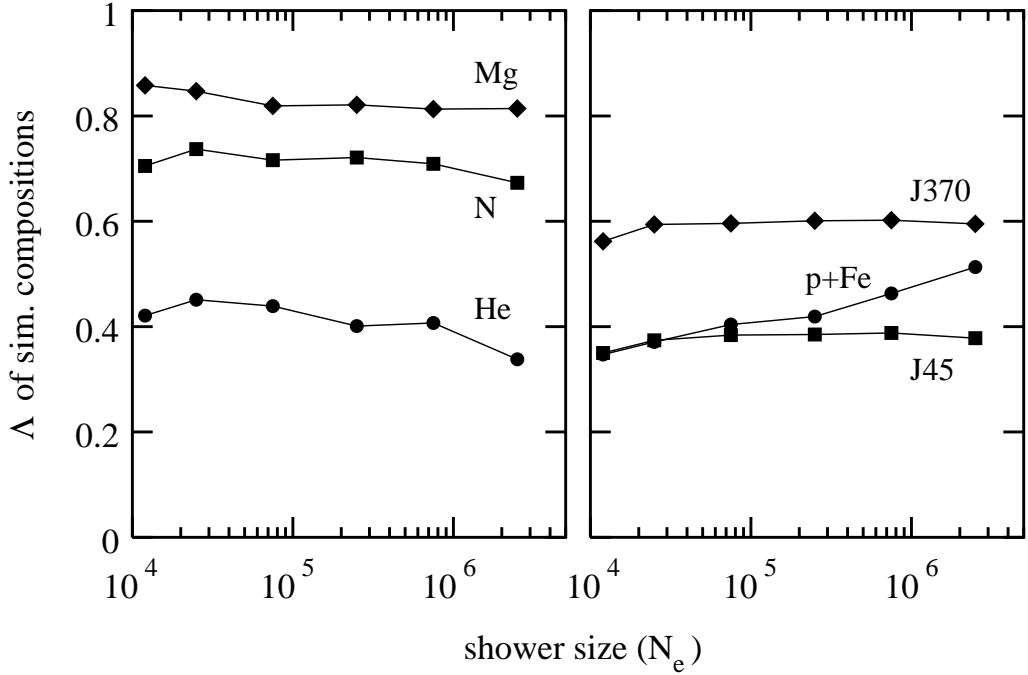

\epsfigure{1.0}{mc-lambda}
\caption[Simulated Lambda values]{Values of $\Lambda$ in simulations
of pure and mixed compositions (assumed constant with a spectrum
slope of $-2.7$). Left: pure Helium ($\ln A/\ln 56=0.344$),
Nitrogen (0.656) and Magnesium (0.790). Right: a mixture of
50\% protons plus 50\% iron ($\langle\ln A\rangle/\ln 56=0.5$)
and compositions like those measured by JACEE \cite{Asakimori-1993ab}
above 45 and 370 TeV ($\langle\ln A\rangle/\ln 56=0.379$
and $0.592$), respectively.}
\label{fig:mc-lambda}
\end{figure}

It turns out that $\Lambda$ is a good measure of $\langle\ln A\rangle$,
in particular for realistic mixed compositions like those from
direct measurements of the JACEE collaboration \cite{Asakimori-1993ab}
above 45 TeV and above 370 TeV. 
In the following, these compositions will be referred to as J45 
and J370, respectively. 
In fact, 
\begin{equation}
\Lambda\approx\langle\ln A\rangle\,/\,\ln 56
\end{equation}
holds with
good accuracy for mixed compositions (see figure~\ref{fig:mc-lambda}),
as long as the composition does not change substantially within
a factor of 2 in energy and the spectrum slope is near $-2.7$.
For this reason we can define
\begin{equation}
\langle\ln \tilde A\rangle = \Lambda \, \ln 56
\end{equation}
to remind of the above relation 
($\langle\ln \tilde A\rangle\approx\langle\ln A\rangle$).

There are two notable deviations visible in 
figure~\ref{fig:mc-lambda}. Pure elements between protons and iron
are all a bit above the corresponding value of $\ln A\,/\,\ln 56$
but slightly falling with shower size as the shower maximum for
proton showers is approaching the HEGRA altitude of 2200~m.
In the extreme case of a composition made up of only protons and iron,
$\Lambda$ is below $\langle\ln A\rangle\,/\,\ln 56$ for the
smaller showers and is rising to the expected value as the
difference in energy of proton and iron showers with the same size $N_e$
changes from a factor of 3 to a factor of 2  
(see table~\ref{tab:E-in-ranges}). For realistic mixed compositions
both effects cancel remarkably well over the whole range of
shower sizes.

For evaluation of experimental $\Lambda$ values a fixed range of
core distances should be preferred where measurements are of
good quality over the whole range of shower sizes.
Because of the CRT electronics saturation when the detector
is hit by more than some 10--20 charged particles (and track
reconstruction starting to deteriorate already before that), 
core distances of less than 80~m were ignored. Note that
for the largest size interval we see some saturation up to
about 100~m core distance. Punch-through electrons are
of no concern for CRT detectors at these large core distances.
Occasional non-shower muons are negligible at even the largest
core distances and small showers \cite{Bernloehr-1996b}.
In the following, results were obtained from data in the
80--280~m core distance range. Reasonable variations of the
distance limits (e.g. 100--250~m) give consistent results.
Only in the size interval $5\mal10^5<N_e<1.5\mal10^6$
there was any noticeable dependence on the lower limit of
the distance range -- which is consistent with a statistical
fluctuation but could be an artefact of detector saturation -- 
and this variation has been included into the
statistical error estimates.

\begin{figure}
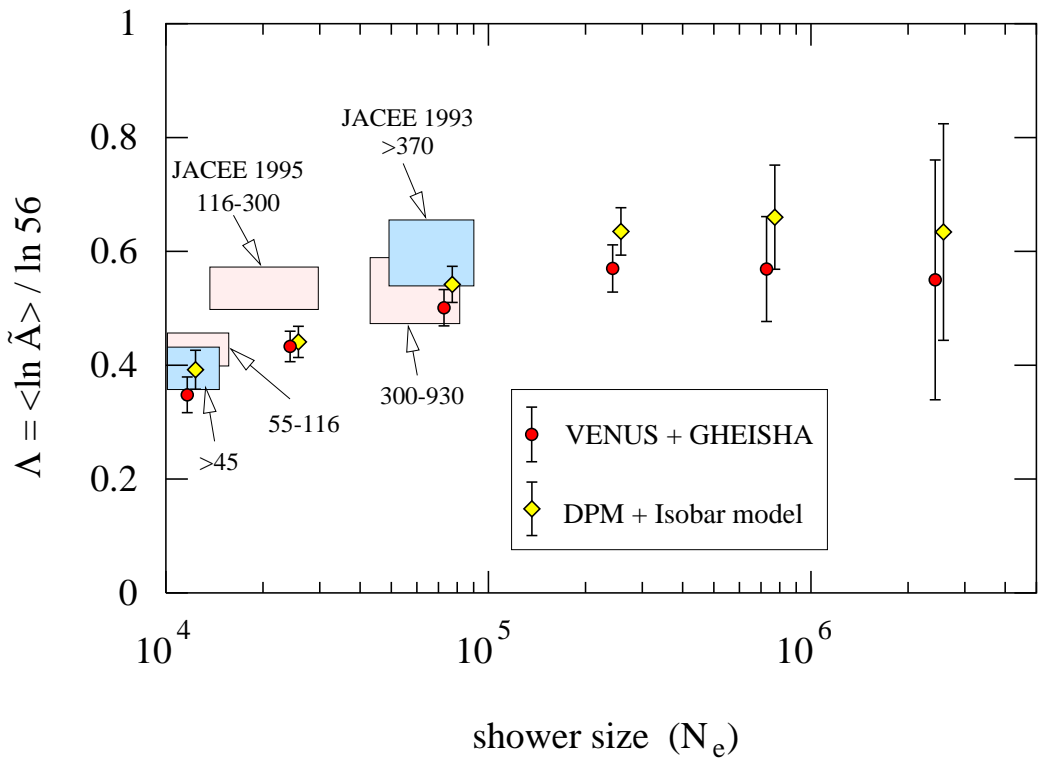

\epsfigure{1.0}{lambda}
\caption[Lambda values of CRT/HEGRA data]{$\Lambda$ values obtained
by comparing CRT/HEGRA data with simulations based on the
VENUS/GHEISHA and DPM/Isobar interaction models. Error bars shown
include only statistical errors of measured data and simulations.
See text for systematic error estimates.
For comparison the $\Lambda$ values corresponding to direct
JACEE composition measurements \cite{Asakimori-1993ab,Tominaga-1995}
are superimposed at the respective {\em typical} shower sizes
(best matching for helium and medium heavy nuclei, see 
table~\ref{tab:E-in-ranges}). The box heights represent quoted
statistical errors.}
\label{fig:Lambda-result}
\end{figure}

\begin{table}
\caption[Lambda values of CRT/HEGRA data]{$\Lambda$ from CRT
and HEGRA data as in figure~\ref{fig:Lambda-result}.}
\label{tab:Lambda-result}
\def\fnm#1{\setcounter{footnote}{0}\footnotemark}
\def\phfnm{\phantom{\setcounter{footnote}{0}\footnotemark}}
\begin{center}
\begin{tabular}{ccc}
 & \multicolumn{1}{c}{$\Lambda$ with} &
  \multicolumn{1}{c}{$\Lambda$ with} \\
 & \multicolumn{1}{c}{VENUS +} &
  \multicolumn{1}{c}{DPM +} \\
$\langle N_e\rangle$ & \multicolumn{1}{c}{GHEISHA} &
  \multicolumn{1}{c}{Isobar} \\
\noalign{\vspace{4pt}}
\hline
\noalign{\vspace{4pt}}
$1.2\mal10^4$ & 0.35 $\pm$ 0.03\fnm{a} & 0.39 $\pm$ 0.03\fnm{a} \\
$2.5\mal10^4$ & 0.43 $\pm$ 0.03\phfnm  & 0.44 $\pm$ 0.03\phfnm \\
$7.5\mal10^4$ & 0.50 $\pm$ 0.03\phfnm  & 0.54 $\pm$ 0.03\phfnm \\
$2.5\mal10^5$ & 0.57 $\pm$ 0.04\phfnm  & 0.63 $\pm$ 0.04\phfnm \\
$7.5\mal10^5$ & 0.57 $\pm$ 0.09\phfnm  & 0.66 $\pm$ 0.09\phfnm \\
$2.5\mal10^6$ & 0.55 $\pm$ 0.21\phfnm  & 0.63 $\pm$ 0.19\phfnm \\
\noalign{\vspace{4pt}}
\hline
\noalign{\vspace{4pt}}
\multicolumn{3}{@{}l}{\fnm{}\footnotesize\parbox[t]{18em}{%
{Statistical errors of data and simulations only. 
See text for systematic errors.}}}
\end{tabular}
\end{center}
\end{table}

Resulting $\Lambda$ values for the data from
figure~\ref{fig:781-786} depend slightly on the interaction model
used for the simulations. Figure~\ref{fig:Lambda-result} and 
table~\ref{tab:Lambda-result} show
results with respect to the VENUS/GHEISHA and the DPM/Isobar
model simulations above/below energies of 80~GeV. For both choices
of models there is good agreement of our results with direct
measurements of JACEE \cite{Asakimori-1993ab,Tominaga-1995}, 
in absolute numbers as well as in the rise of $\Lambda$ with shower size
or energy. Statistical errors for $\Lambda$ were checked by
simulations given the measured numbers of muons in each
distance bin.

In addition to the statistical errors shown in figure~\ref{fig:Lambda-result}
there are several sources of systematic errors.
These are either independent of the shower size, like the
calibration of detector alignments (0.06)
or only very weakly depending on shower size, like the relation
of shower sizes in equation~\ref{eq:Nh/Ne} (0.03 overall plus
$0.015\log_{10}(N_e/3\mal10^4)$ size dependent).
The systematic error related to the interaction model is estimated
as 0.05 (see figure~\ref{fig:Lambda-result}) and might depend
weakly on shower size. The overall systematic error of $\Lambda$
is thus estimated as 0.08 (common to all size intervals) and
the error of $d\Lambda/d\log_{10}(N_e/3\mal10^4)$ as 0.025.
Considering these errors, the agreement with JACEE is remarkably good.


\section{Interpretation}

Any interpretation of composition measurements should also
take the available data on the overall cosmic-ray flux
into account. A particularly simple model of the cosmic-ray
composition is to assume that there is only one universal type
of Galactic cosmic-ray sources and the acceleration time scales as well
as the propagation are then only functions of the rigidity 
$R=pc/Ze$ ($p$ being the momentum).
It is well known that such a simple model cannot
be matched very well to the older Akeno measurements \cite{Nagano-1984}
where the knee appears almost as a single kink in the flux spectrum.
The newer Tibet array measurements \cite{Amenomori-1996a}
with a gradual steepening of the spectrum are much closer
to expectations for a minimal model but still the steepening
of the spectrum has to happen within less than about 
half a decade of rigidity.

\begin{figure}
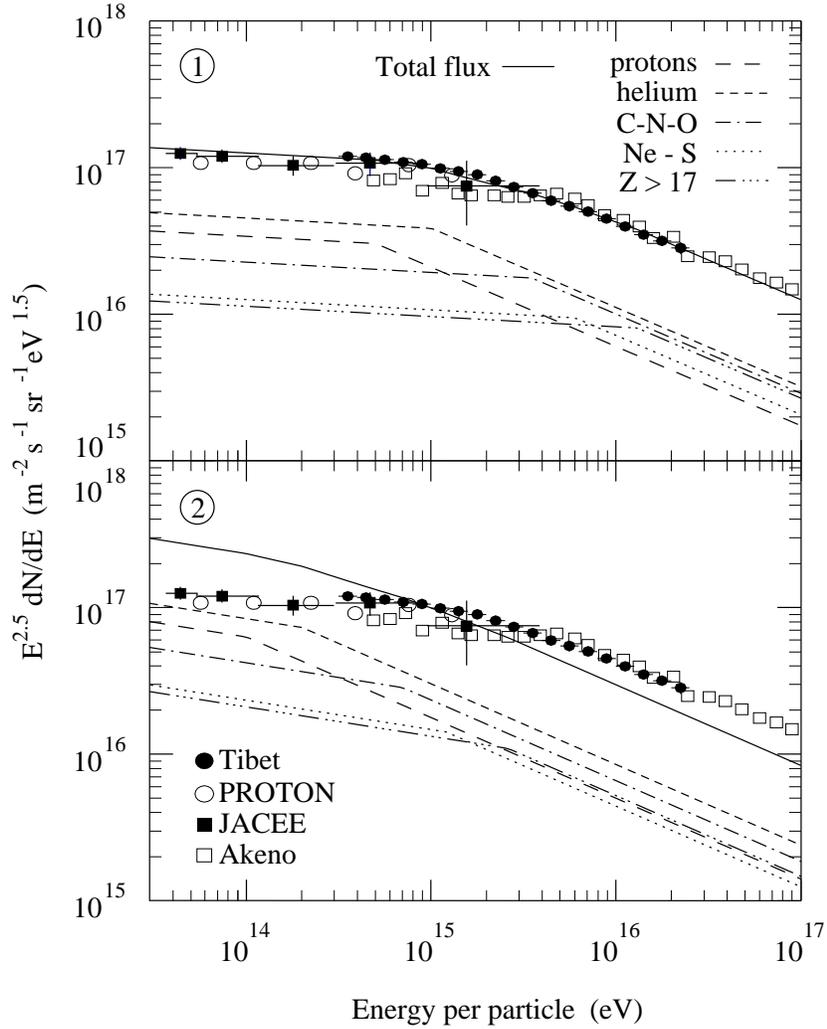

\epsfigure{0.8}{flux_models_1+2}
\caption[Composition models 1 and 2]{Fluxes scaled by energy $E^{2.5}$ of 
model compositions with a simple knee at fixed rigidity. 
Model 1: slope -2.57 below the knee rigidity of 500 TV and -3.04 above. 
Model 2: slope -2.7 below 100 TV and -3.05 above. Models are normalized
to Tibet array measurements (for their HD1 model)
\cite{Amenomori-1996a} at 1~PeV.
Also shown for comparison are flux measurements of the
Proton \cite{Grigorov-1971}, JACEE \cite{Tominaga-1995} and
Akeno \cite{Nagano-1984} experiments.
Note that differences of the order of 30\% between absolute flux 
scales by different experiments are consistent with their estimated 
energy scale errors.}
\label{fig:models-1+2}
\end{figure}

A minimal composition model would be to assume a sudden kink
of each component at a given rigidity.
Such a simple model is consistent with our composition data alone
(for $R_{\rm knee}\approx100$~TV, see model 2 in
figure~\ref{fig:composition-models}) or with the flux data
of JACEE and the Tibet array \cite{Amenomori-1996a} alone 
(for $R_{\rm knee}\approx500$~TV, see model 1 in 
figure~\ref{fig:models-1+2}) but 
could not be tuned to be consistent with both our composition data
and the available flux data. This inconsistency is even more
severe if the Akeno flux spectrum \cite{Nagano-1984} is assumed.

\begin{figure}
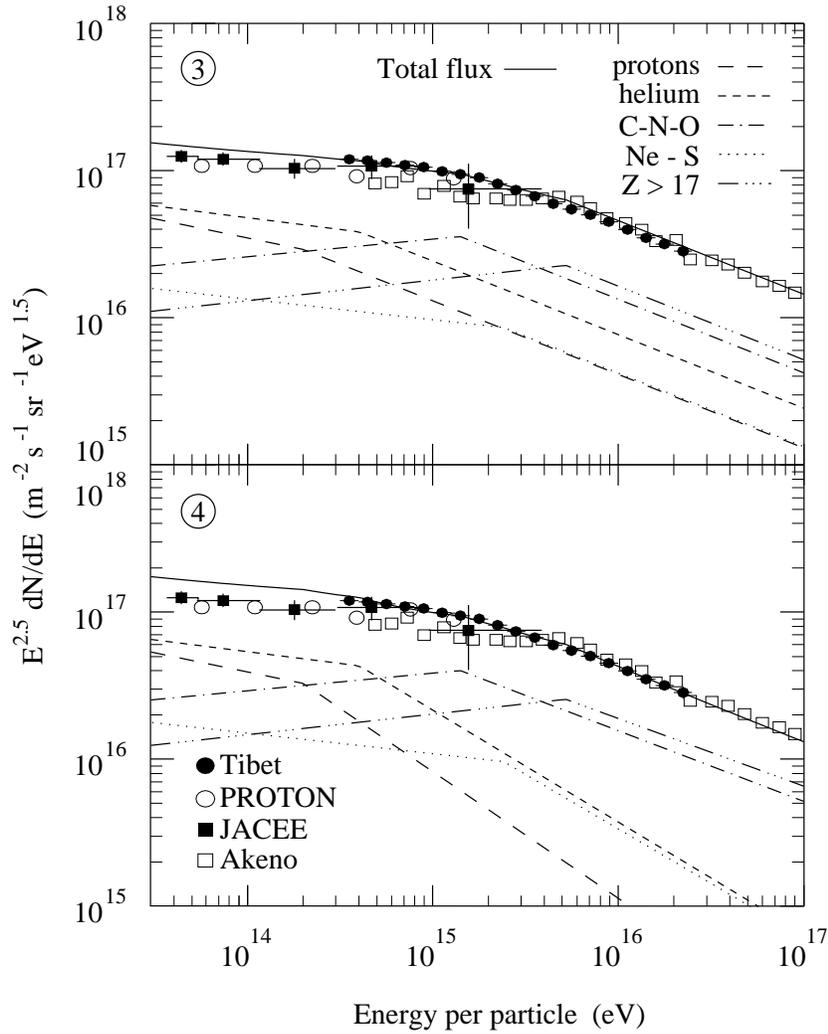

\epsfigure{0.8}{flux_models_3+4}
\caption[Composition models 3 and 4]{Fluxes of model compositions 
extrapolating JACEE measurements with an additional knee at a fixed
rigidity of 200~TV. Model 3: slope changes to -3.01 above the knee. 
Model 4: slope steepens by 0.6 at the knee.}
\label{fig:models-3+4}
\end{figure}

Such a simple model is also inconsistent with JACEE composition
measurements finding a harder spectrum for the CNO and iron groups
of nuclei than for protons and helium near 100~TeV. Although it is
usually believed that there is a transition to another dominant
source type at the knee, neither are the JACEE data precise enough
nor is our measurement of $\langle\ln A\rangle$ sufficient to
conclude that such a transition is already seen. If there is
a cutoff in the spectrum of one source component (presumably supernova
remnants) and a transition to an entirely different one,
it is understandable that cutoff and transition should 
happen at about the same energy but it remains very puzzling that both
components should add up to form the knee. Why should the
second component not extend with its $E^{-3}$ spectrum to lower
energies but just set on at the same energy and with the same flux 
where the first component cuts off.
Assuming there is a transition at the knee, then the second component
has a heavier composition but -- based on the JACEE data and our
$\langle\ln A\rangle$ -- not only consisting of iron group nuclei.

\begin{figure}
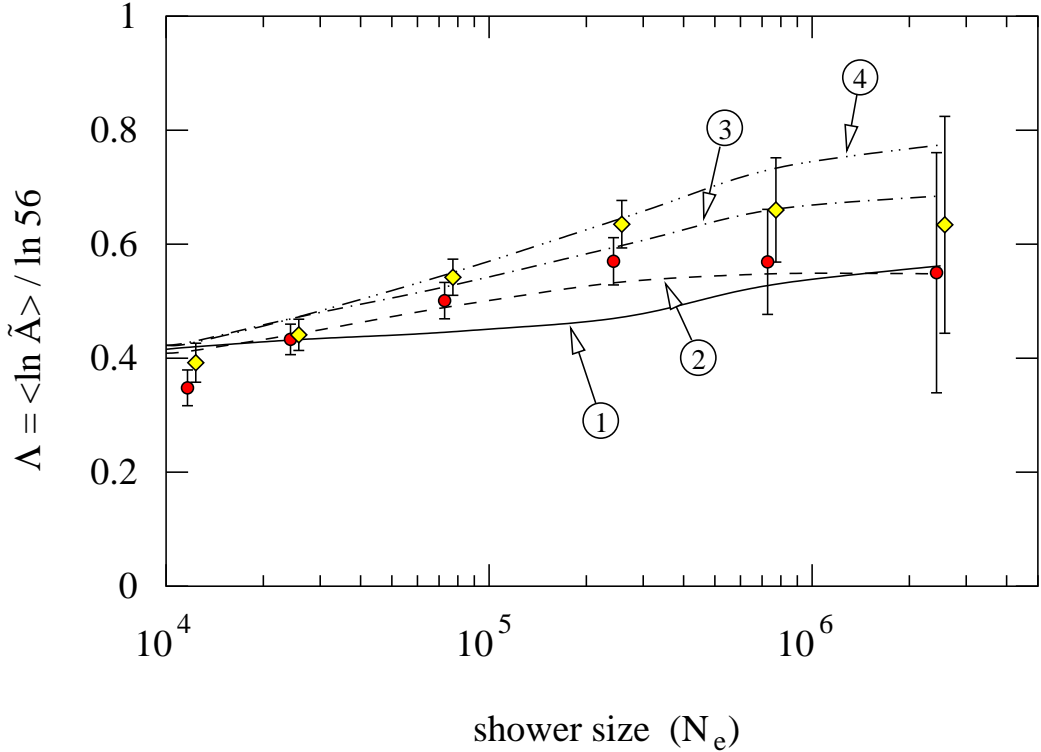

\epsfigure{1.0}{lambda-n3+model}
\caption[Lambda values corresponding to models]{Values of $\Lambda$
corresponding to composition models shown in figures
\ref{fig:models-1+2} and \ref{fig:models-3+4}
compared to CRT/HEGRA results for both choices of interaction models
as in figure~\ref{fig:Lambda-result}.}
\label{fig:composition-models}
\end{figure}

Without firm physical models of the cutoff of a first
and the onset of a second component
there are too many free parameters for reasonable checks.
It is perhaps more illustrating to see how an extrapolation of
JACEE measurements combined with a rigidity dependent knee
compares with composition and flux data. Starting with
a composition in the 55--116 TeV energy range
and spectral indices of the five groups of nuclei  as measured by JACEE
\cite{Tominaga-1995}, there are just two parameters in such models:
the rigidity of the knee for each element and either the
slope change at the knee or a common spectral index above the knee.
Figure~\ref{fig:models-3+4} shows such models fitting the Tibet array
\cite{Amenomori-1996a} and other flux measurements.
The knee rigidity for such models is about 200~TV.
Figure~\ref{fig:composition-models} shows that both such models
are consistent with our composition measurements.

\section{Conclusions}

With the measurements of muon track angles in extensive air
showers it is possible to obtain an `average mass' parameter $\Lambda$
of the cosmic-ray chemical composition in the $10^{14}$--$10^{16}$~eV
energy range with $\Lambda\approx\langle\ln A\rangle/\ln 56$.
The bias of air-shower arrays in favour of showers induced by
light primaries is cancelled to a very good approximation by the
larger number of muons in showers of heavy nuclei.

The analysis of the median radial angles of muons in the
CRT/HEGRA data set shows an increase of $\langle\ln A\rangle$
in the $10^{14}$--$10^{15}$~eV which agrees very well with
the highest energy direct measurements made by the JACEE collaboration.
A combined analysis of the CRT/HEGRA composition data and
cosmic-ray flux data of several experiments shows that a
{\em minimal} composition model with a simple knee at a fixed rigidity
can only be matched to either the composition or the flux data
but not simultaneously to both. Although it may be too early to
conclude that the transition to a second type of Galactic cosmic-ray
sources is seen, the composition of cosmic-rays from a possible
second type of sources should be heavier than what is measured
well below the knee but not made up entirely of heavy nuclei.
At the present level of accuracy of both composition and flux
measurements an extrapolation of direct composition
measurements with the additional constraint of a kink at a
fixed rigidity of about 200~TV can describe the flux and composition
measurements. 

With the additional available data covering about half a year of
combined CRT and HEGRA
data taking (a four-fold of present statistics) the statistical errors
even in our largest shower size interval would be well below present
systematic uncertainties of the shower simulations. An extension to
substantially larger shower sizes, however, would be difficult with
the present installation because both CRTs and HEGRA would be
affected by detector saturation.


\section*{Acknowledgements}

We wish to thank the HEGRA collaboration very much 
for supporting the installation and operation of the CRT detectors at the
HEGRA site and for providing the array data used in our analysis.
It is a pleasure to thank also the authors of the CORSIKA shower simulation
program, in particular J.~Knapp and D.~Heck. 
We thank the authorities of the 
Instituto de Astrof{\'\i}sica de Canarias (IAC) and the
Roque de los Muchachos Observatory (ORM) for the excellent working 
conditions and for allowing
measurements with CRT detectors outside of the HEGRA area.
We gratefully acknowledge the technical support by the
mechanical and electronic workshops of the MPIK.



\end{document}